\documentclass[aps,prb,reprint,superscriptaddress]{revtex4-2}
\usepackage{placeins}
\usepackage{dcolumn}
\usepackage{graphicx}
\usepackage{amsmath,amssymb,bm}
\usepackage{xr, xfrac}

\usepackage{color}
\usepackage{multirow}
\usepackage{textcomp}
\usepackage{ragged2e}

\usepackage{multibib}
\newcites{SI}{Supplementary Information References}

\makeatletter
\renewcommand{\@makecaption}[2]{%
  \par\small
  \begingroup
  \setlength{\parindent}{0pt}
  \justifying
  \noindent\textbf{#1.} #2\par
  \endgroup}
\makeatother
\usepackage[colorlinks=true, linkcolor=black, urlcolor=blue, citecolor=blue]{hyperref}

\begin{document}

\title{Probing tunable Kerr nonlinearity in graphene Josephson junctions}

\author{Priyanka Samanta}
\email{samantapriyanka814@gmail.com}
\affiliation{Department of Condensed Matter Physics and Materials Science, Tata Institute of Fundamental Research, Mumbai 400005, India}

\author{Joydip Sarkar}
\affiliation{Department of Condensed Matter Physics and Materials Science, Tata Institute of Fundamental Research, Mumbai 400005, India}

\author{Ashish Abhraham Samuel}
\affiliation{Department of Condensed Matter Physics and Materials Science, Tata Institute of Fundamental Research, Mumbai 400005, India}
\affiliation{Centre for Advanced Studies in Electronic Science and Technology, University of Hyderabad, Hyderabad 500046, India}

\author{Madhavi Chand}
\affiliation{Department of Condensed Matter Physics and Materials Science, Tata Institute of Fundamental Research, Mumbai 400005, India}

\author{Kenji Watanabe}
\affiliation{Research Center for Functional Materials,\\ National Institute for Materials Science, Tsukuba 305-0044, Japan}

\author{Takashi Taniguchi}
\affiliation{International Center for Materials Nanoarchitectonics, National Institute for Materials Science, Tsukuba 305-0044, Japan}

\author{Mandar M. Deshmukh}
\email{deshmukh@tifr.res.in}
\affiliation{Department of Condensed Matter Physics and Materials Science, Tata Institute of Fundamental Research, Mumbai 400005, India}

\begin{abstract}
\vspace{0.05in}

    Josephson junction (JJ) is a key nonlinear element in superconducting devices such as qubits, amplifiers, and bolometers. Recently, gate-tunable JJs based on graphene and semiconductors have gained interest due to their rich Andreev physics and wide applications in circuit quantum electrodynamics devices. In addition to gate tunability, it offers many advantages over conventional JJs, such as exceptional thermal properties for bolometric sensors, magnetic-field compatibility, and operability at elevated temperatures above $1$ K. Like conventional Al-AlO$_\mathrm{x}$-Al JJs, graphene JJs also act as nonlinear inductors, and at their heart lies the Kerr nonlinearity. Additionally, in graphene JJs, the nonlinearity is tunable via external knobs in a single device. However, a detailed exploration of the tunable Kerr nonlinearity in graphene JJs has never been performed. In this work, we study the dependence of the Kerr nonlinearity on gate voltage, temperature, and DC bias -- an interesting knob that has been less explored. Using these parameters, we show that the magnitude of the Kerr coefficient can be tuned over a wide range, from $\sim 300$ kHz to $1.2$ MHz. Our work will be a valuable resource for further understanding of the nonlinearity in graphene JJs and for the design of next-generation amplifiers and sensors, with enhanced performance.

\end{abstract}
\maketitle

\vspace{2em}

Josephson junctions (JJs) constitute the fundamental nonlinear elements of superconducting circuits and form the basis of many modern quantum technologies~\cite{ clarke2008, devoret2013}. A JJ supports a dissipationless supercurrent described by the current-phase relation (CPR) $I_\mathrm{s} = I_\mathrm{c} \sin{\phi}$, where $I_\mathrm{c}$ is the critical current and $\phi$ is the superconducting phase difference across the junction for conventional superconductor-insulator-superconductor (SIS) JJs~\cite{josephson1962, tinkham1996, likharev1979, Golubov2004CPR}. From the Josephson relations it can be shown that the JJs have an effective inductance $L_\mathrm{J}(\phi)=\frac{\Phi_0}{2\pi} \left( \frac{\partial I_\mathrm{s}}{\partial\phi}\right)^{-1}=\frac{\Phi_0}{2\pi I_\mathrm{c} \cos{\phi}}$, known as the Josephson inductance where $\Phi_0$ is the magnetic flux quanta~\cite{tinkham1996, likharev1979}. The phase dependence of this inductance makes JJs intrinsically nonlinear circuit elements, enabling the realization of tunable microwave resonators and nonlinear superconducting devices~\cite{wallraff2004, siddiqi2004, yamamoto2008, castellanos2007}. Such nonlinearities help realize a wide range of applications, including superconducting qubits~\cite{koch2007, clarke2008, devoret2013}, parametric amplifiers~\cite{yamamoto2008, castellanos2007, macklin2015, bergeal2010, hao2026}, and quantum-limited detectors~\cite{zmuidzinas2012, day2003}. In particular, the ability to engineer strong yet nearly dissipation-less nonlinearities has made Josephson circuits a powerful platform for circuit quantum electrodynamics (cQED), enabling strong light-matter coupling and high-fidelity quantum control of microwave photons~\cite{wallraff2004, blais2004, blais2021}. Recent experiments in hybrid and two-dimensional material platforms, including graphene and semiconductor-based JJs, have demonstrated strongly nonlinear behavior and the emergence of unconventional CPR with higher-harmonic contributions, such as $\sin{2 \phi}$, and even fractional $4\pi$ periodic components~\cite{english2016, messelot2024, GeJunction2025, WTe2JJ2023}, along with the added knob of electrostatic gate tunability. These developments highlight the growing ability to engineer and probe nonlinear Josephson elements on emerging material platforms~\cite{Sarkar2026vdWJJ, Generalov2024}.

\begin{figure*}[ht!]
    \centering
    \includegraphics[width=0.9\textwidth]{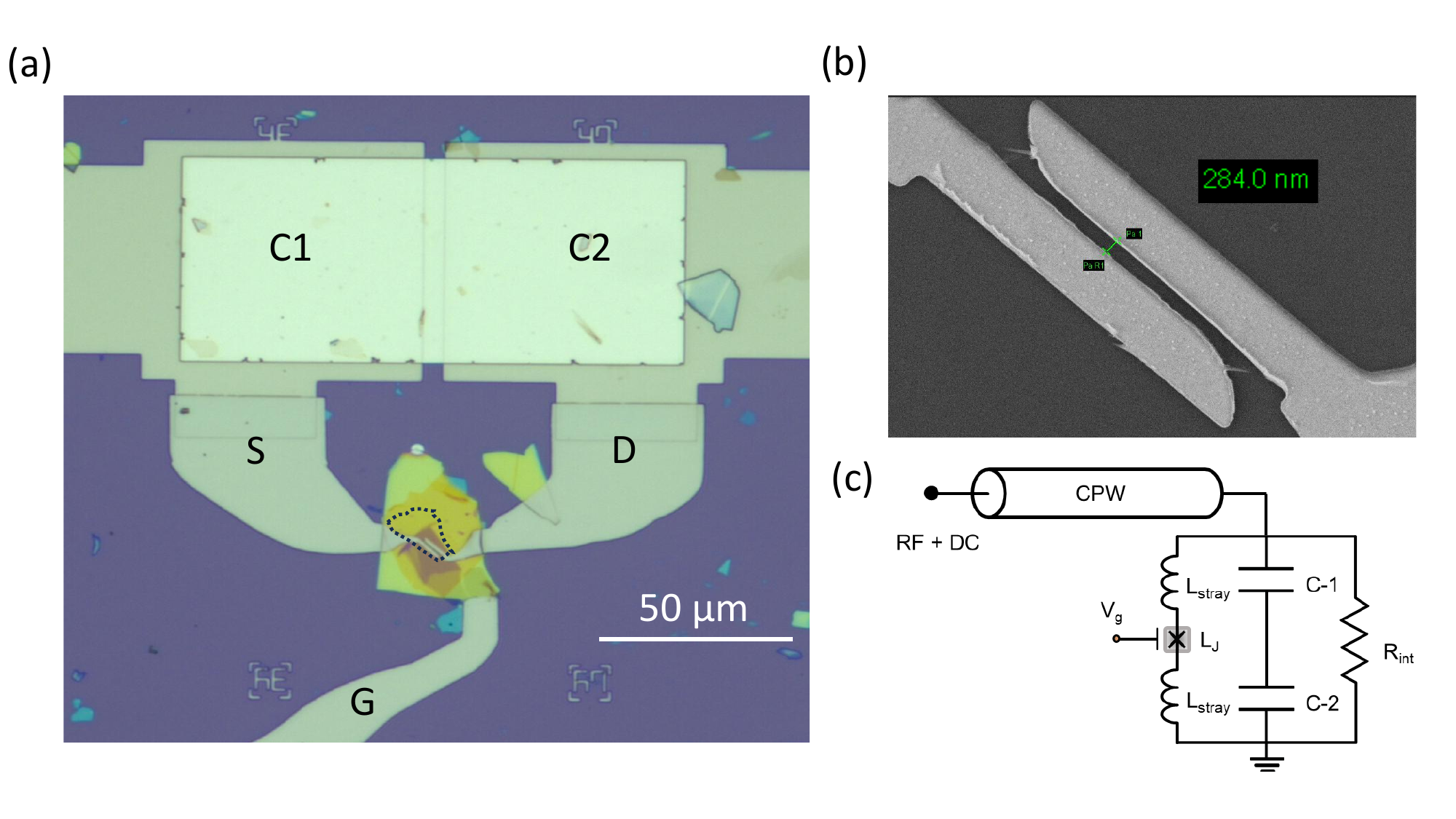}
    \caption{({\textbf a}) The optical micrograph of the graphene Josephson junction device with a local graphite back gate. The marked parts correspond to the source (S), drain (D), and gate (G) electrodes, which were deposited using MoRe, a superconducting alloy. We also make two capacitors (C1 and C2) in series, which are in parallel to our Josephson junction.
    ({\textbf b}) The scanning electron micrograph shows a zoomed-in view of the JJ. The distance between the $2$ electrodes is measured to be $\sim 280$ nm. 
    ({\textbf c}) The effective microwave circuit for our device can be modeled as a parallel LCR circuit directly coupled to a CPW transmission line for reflection measurements.}
    \label{fig:f1}
\end{figure*}
For SIS JJs, the Kerr nonlinearity arises from the cosine potential of the junction, with the potential energy given by $U(\phi)=-E_\mathrm{J}\cos{\phi}$ in the absence of a DC bias, where $E_\mathrm{J}$ is the Josephson energy~\cite{josephson1962, tinkham1996, likharev1979, Golubov2004CPR}. Expanding the potential about its minimum, the quadratic term describes a harmonic oscillator, while the quartic term captures the leading-order anharmonic correction~\cite{blais2004, blais2021}. For graphene JJs, the potential takes a different form, given by a distinct CPR arising from the Andreev bound states, as we discuss in the following section. However, the nature of the nonlinearity and its general properties are similar to those of SIS JJs but with added knobs for tunability in a single device. 
 
Kerr nonlinearities in superconducting resonators and JJ arrays have been investigated both theoretically and experimentally, revealing photon-photon interactions, nonlinear microwave dynamics, and parametric amplification processes~\cite{krupko2018, white2015}. 
Recent efforts on superconductor-semiconductor JJs have enabled a variety of nonlinear cQED devices, including gatemon qubits~\cite{larsen2015, casparis2018, wang2019, larsen2020_parity, balgley2025}, quantum-limited parametric amplifiers~\cite{sarkar2022, butseraen2022, phan2023, hao2024}, and bolometric sensors~\cite{lee2020, sarkar2025}. 

As discussed earlier, the nonlinear inductive response of a JJ arises from the anharmonicity of the junction potential. We now discuss the nature of nonlinearity in SNS JJs, e.g., graphene JJs, where the CPR is governed by Andreev bound states. For a single transport channel with transmission probability $\tau$, the ground-state energy of the Andreev bound state is given by $U(\phi) = -\Delta_0 \sqrt{1 - \tau \sin^2(\phi/2)}$ and the resultant CPR: $I_\mathrm{s}(\phi) = \frac{\pi \Delta_0}{2 e R_\mathrm{n}} \, \frac{\sin \phi}{\sqrt{1 - \tau \sin^2(\phi/2)}}$, where $I_\mathrm{s}$ is the supercurrent, $\phi$ is the phase drop across the JJ, $\Delta_0$ is the induced superconducting energy gap, $e$ is the electronic charge, and $R_\mathrm{n}$ is the normal state resistance~\cite{titov2006, Kringhoj2018Anharmonicity}. The Kerr coefficient is obtained by expanding the potential around its minimum ($\phi_\mathrm{min} = 0$) and extracting the nonlinear coefficients. Writing the Taylor series expansion
$U(\phi) \approx -E_\mathrm{J}(\frac{1}{2!} c_2\phi^2 +\frac{1}{3!} c_3\phi^3 + \frac{1}{4!} c_4\phi^4 + \cdots),$ where the dimensionless coefficients are defined as
$c_n = \frac{1}{E_\mathrm{J}} \frac{d^n U}{d\phi^n}\Big|_{\phi_\mathrm{min}}$ and $E_\mathrm{J}=\frac{d^2 U}{d\phi^2}\Big|_{\phi_\mathrm{min}}$.
Evaluating the expansion of $U(\phi)$ for SNS JJ yields, $c_2 = 1$, $c_3 = 0$, and $c_4 = -1+\frac{3\tau}{4}.$
For a single JJ-based resonator, the Kerr coefficient is then primarily set by $c_4$ and the inductive participation ratio $p=L_\mathrm{J}/(L_\mathrm{J}+L_\mathrm{stray})$ representing the contribution of the JJ's inductance $L_\mathrm{J}$ over other parasitic sources of inductance $L_\mathrm{stray}$, such that $K \propto (c_4-\frac{5c_3^2}{3c_2})\, p^3  = -(1-\frac{3\tau}{4})\,p^3$, see Supplementary Information section III for the derivation of the Kerr coefficient expression from nonlinear Duffing oscillator's equation of motion~\cite{hao2024, Zhou2014JPA}.
It can be noted that the Kerr nonlinearity depends strongly on the transparency of the junction, and in SNS JJs with high-transparency 
($\tau\approx1$), the nonlinearity is reduced to $25\%$ to that of tunnel JJs ($\tau\approx0$).

\begin{figure*}[ht!]
    \includegraphics[width=\textwidth]{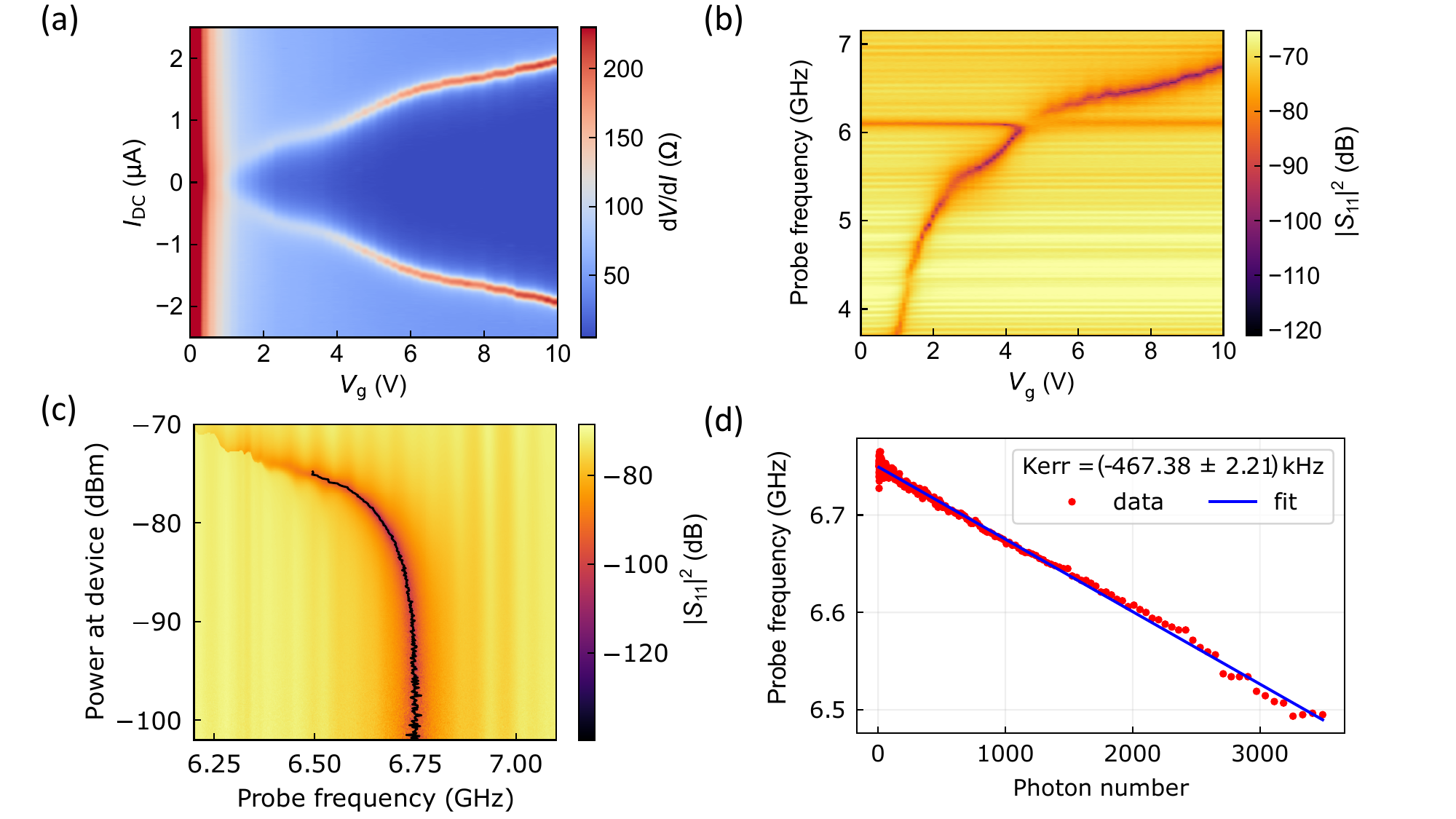}
    \caption{({\textbf a}) A DC pseudo-$4$-probe measurement showing the effect of gate voltage ($V_\mathrm{g}$) and DC bias ($I_\mathrm{DC}$) on the differential resistance (d$V$/d$I$) of the graphene JJ. 
    ({\textbf b}) A microwave measurement showing the dependence of the resonance frequency on the applied gate voltage ($V_\mathrm{g}$), where the color scale represents the magnitude of the reflection coefficient ($|S_\mathrm{11}|^2$). 
    ({\textbf c}) At $V_\mathrm{g} = 10$ V and $ T = 20$ mK, the device's power-dependent response is characterized. The resonant frequency corresponding to each applied microwave probe power is extracted by tracking the dip in the $|S_\mathrm{11}|^2$ signal. These extracted resonant frequencies are subsequently used to estimate the average photon number associated with each measurement condition.
    ({\textbf d}) A linear regression of the resonant frequency as a function of the average photon number, corresponding to the data presented in panel (c), is performed to extract the Kerr coefficient.
    }
    \label{fig:f2}
\end{figure*}

The magnitude of the Kerr coefficient depends strongly on the junction type and circuit parameters, like the Josephson energy $E_\mathrm{J}$, charging energy $E_\mathrm{C}$, and participation ratio $p$~\cite{Bourassa2012, Boutin2017}. For instance, in transmon qubits based on SIS JJs, the Kerr coefficient is typically set by the charging energy $K \approx -E_\mathrm{C}/\hbar$, where $\hbar$ is the reduced Planck's constant, yielding values on the order of $100$--$300$ MHz~\cite{koch2007, schreier2008}. In contrast, weakly nonlinear microwave resonators incorporating JJ or SQUID arrays exhibit relatively smaller Kerr nonlinearity, typically in the range of $10$ kHz to a few MHz, enabling their use in parametric amplification and quantum-limited detection~\cite{yamamoto2008, castellanos2007, macklin2015}. For JPAs, the Kerr coefficient K/2$\pi$ is desired to be in the range of $\sim$ $1$ to $100$ kHz, as devices with weak nonlinearity require high pump powers to operate, while very high nonlinearity lowers the compression point in JPAs~\cite{zapata2024}. In practice, having control knobs to tune the Kerr coefficient in a single device is a very demanding feature and is less explored in conventional SIS JJ-based devices. Recent experiments in hybrid superconducting 2D material systems have demonstrated gate-tunable CPR~\cite{Haller2022,messelot2024, Huang2023Shapiro}, which governs the degree of nonlinearity and is crucial for designing superconducting quantum circuits, including protected qubits, parametric amplifiers, and nonlinear single-photon sensors.

In this work, we have studied the Kerr nonlinearity in graphene JJs. The Kerr coefficient depends on the junction transparency and can be tuned by the applied gate voltage in SNS JJs~\cite{Jung2025ABSthermal}. We have studied the dependence of Kerr coefficient on gate voltage, DC bias, and temperature. The DC bias serves as an important knob for effectively reshaping the junction's potential. As we will discuss, our device geometry allows us to bias the JJ with DC and simultaneously probe its microwave response. 
The DC bias can be a powerful knob for engineering a Kerr-free cubic nonlinearity in the potential ($c_3\neq0$, $c_4\approx0$), while still enabling nonlinear mixing to have amplification with enhanced dynamic range similar to asymmetric SQUID-based amplifiers~\cite{Frattini2017ThreeWave, Frattini2018SNAIL}.

We now discuss our device geometry. The weak link in our JJ is realized using a few-layer graphene encapsulated with two hexagonal boron nitride (hBN) flakes. The superconducting electrodes were made by edge contacts by sputtering a type-II superconducting alloy, molybdenum rhenium (MoRe). Thus, our superconductor-normal conductor-superconductor JJ is planar. Below our graphene JJ, we have a thick graphite flake that acts as a local electrostatic gate. Figure \ref{fig:f1}a shows the optical micrograph of the JJ with the hBN-graphene-hBN-graphite
stack. The gap between the two JJ fingers as seen in the zoomed SEM image in Figure \ref{fig:f1}b is $\sim280$ nm. Before making the JJ, the stack is initially dropped onto a coplanar waveguide (CPW) with a characteristic impedance of $50$ $\Omega$, patterned with standard e-beam lithography on a SiO$_2$/Si substrate. In this CPW, we also fabricate two MoRe-Al$_2$O$_3$-Al parallel-plate capacitors in parallel with our graphene JJ, forming a parallel LC resonator. The full device schematic, including the CPW is shown in Supplementary figure 1. More details on the device fabrication are provided in the Supplementary Information section I. 

\begin{figure*}[ht!]
    \centering
    \includegraphics[width=1\textwidth]{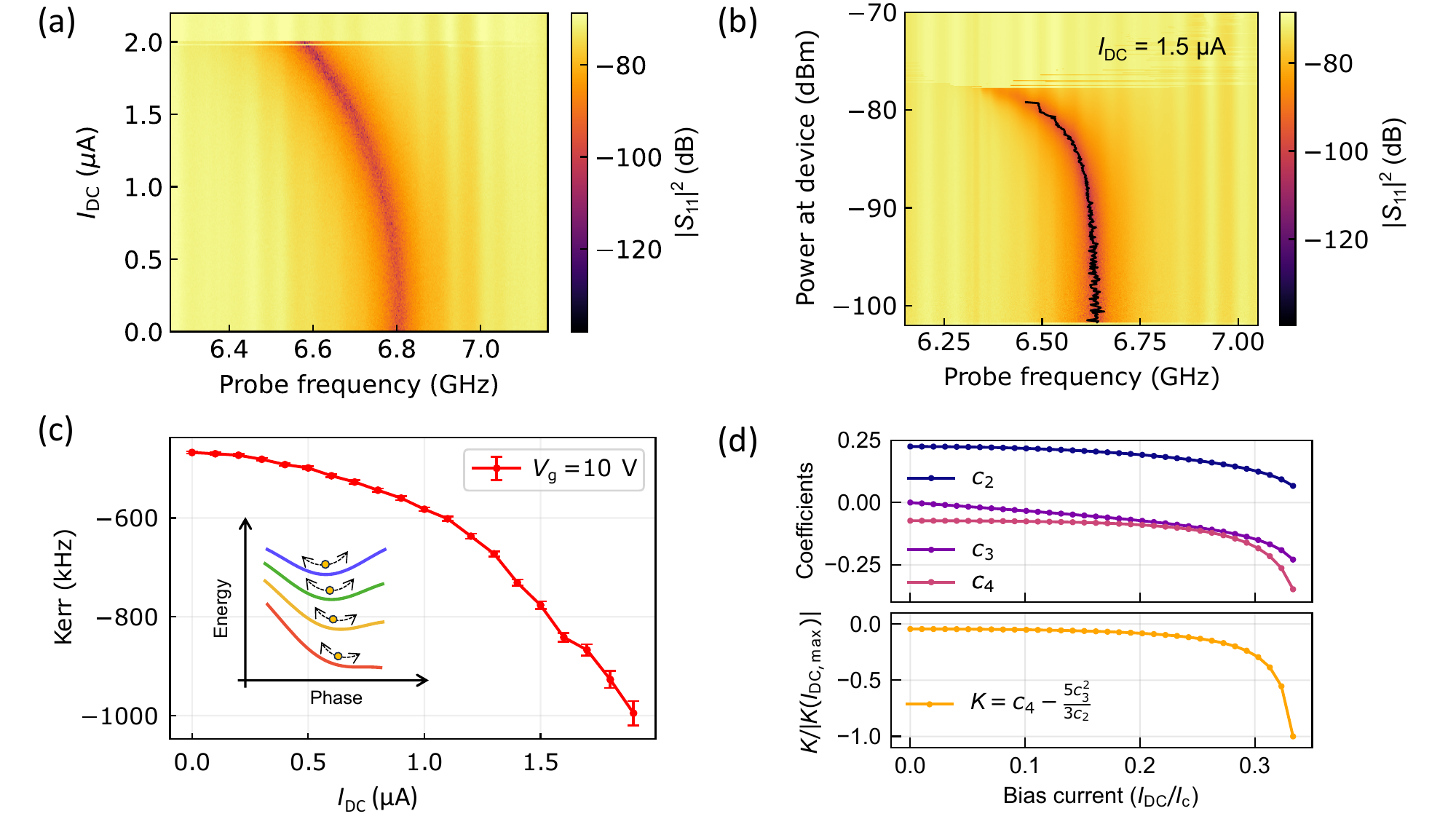}
    \caption{
    ({\textbf a}) The $S_\mathrm{11}$ magnitude response at a gate voltage of $10$ V and a temperature of $20$ mK is measured as a function of the applied DC bias in the low-power linear regime of the probe signal. At higher bias currents, the resonance feature is suppressed and eventually vanishes, indicating a switching of the junction from the superconducting to the normal state. 
    ({\textbf b}) The power-dependent response of the device is measured at different DC bias (here $1.5$ $\mathrm{\mu}$A) and a gate voltage of $10$ V at $20$ mK. The resonant frequency corresponding to each applied microwave probe power is extracted by tracking the dip in the magnitude of the reflection coefficient ($|S_\mathrm{11}|^2$).
    ({\textbf c}) The Kerr coefficient is extracted at each applied DC bias by fitting procedures performed at a fixed gate voltage of 10 V and a temperature of 20 mK. Inset, a schematic illustrating how an increase in the applied DC bias enhances the tilt of the washboard potential in the RCSJ model, thereby reshaping the effective potential experienced by the phase particle. The blue-to-red colors indicate an increase in DC bias strength.
    ({\textbf d}) The analytically calculated bias-current dependence of the nonlinear coefficients is plotted. The coefficients in arbitrary units are calculated by expanding the potential around the minima for an SNS junction under DC bias, with an effective transparency of $\tau=0.9$.
    }
    \label{fig:f3}
\end{figure*}

Before microwave measurements, we perform DC characterization of our graphene JJ to verify its superconductivity, determine its electrostatic doping tunability, and measure its switching current. 
The differential resistance (d$V$/d$I$) map in the pseudo-four-probe configuration is shown in Figure \ref{fig:f2}a. 
There, we measure d$V$/d$I$ by sweeping the gate voltage applied and the DC bias at a temperature of $20$ mK in a dilution refrigerator. 
The dark blue region corresponds to the superconducting state with zero differential resistance, and the red boundaries of that region mark the superconducting-to-normal transition. With higher gate voltages, the critical current increases up to $2$ $\mathrm{\mu}$A at V$_\mathrm{g}$ = $10$ V. At V$_\mathrm {g}$ = $0$ V, we find the charge-neutrality point of graphene, and thus the critical current is zero.
In the microwave measurement, we measure the reflected signal $S_\mathrm{11}$ from the device, as shown in Figure \ref{fig:f1}c. The input signal is attenuated at different temperature stages in the dilution refrigerator to reduce thermal noise before reaching the device. Whereas, the output signal is amplified at the $4$ K stage to improve the signal-to-noise ratio. More details on the measurement setup are provided in the Supplementary Information section II. In Figure \ref{fig:f2}b, the log magnitude of the $S_\mathrm{11}$ signal is plotted. The dispersive feature in the plot corresponds to the tunable resonance frequency controlled by the gate voltage.
The mapping between the DC pseudo four-probe resistance map and the resonance frequency tunability can be understood as follows: the higher the gate voltage $V_\mathrm{g}$, the higher the switching current $I_\mathrm{c}$, and the lower the JJ inductance $L_\mathrm{J}$, since they are inversely proportional. So, the resonance frequency increases with gating.
We observe a circuit resonance at $\sim 6.1$ GHz that remains unchanged with gating. In our device, at resonance, we see a $2 \pi$ change in $S_\mathrm{11}$ phase response, suggesting an over-coupled state of the device where the losses are dominated by the external coupling due to a directly coupled architecture in our device, which sets the loaded quality factor around $10$.

\begin{figure*}[ht!]
    \centering
    \includegraphics[width=\textwidth]{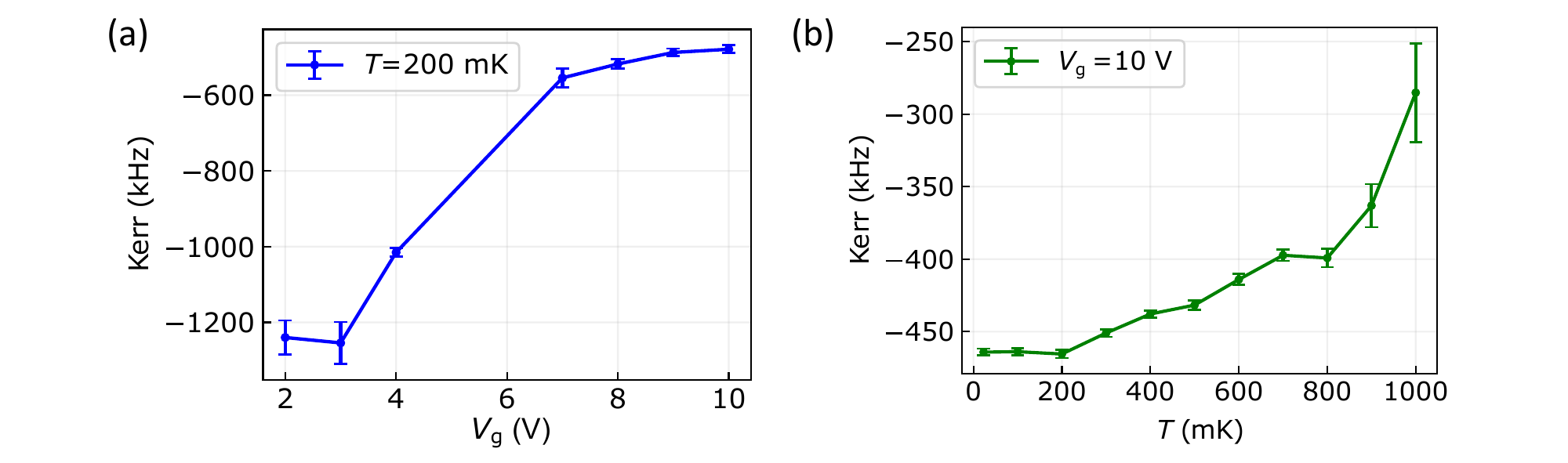}
    \caption{
    (\textbf{a}) The Kerr coefficient is extracted at different gate voltages, at a fixed temperature of $200$ mK.
    (\textbf{b}) The Kerr coefficient is extracted as a function of temperature at a fixed gate voltage of $10$ V. In both cases, the measurements are performed without a DC bias ($I_\mathrm{DC}=0~\mu$A).
    }
    \label{fig:f4}
\end{figure*}

The device's nonlinear response is characterized by fixing the gate voltage and temperature ($V_\mathrm{g}$ = $10$ V, $T$ = $20$ mK) and increasing the applied microwave probe power while monitoring the resonance frequency. The resulting power-dependent frequency shift is shown in Figure \ref{fig:f2}c. Two distinct regimes are observed. At low drive powers, the resonance frequency remains essentially unchanged, indicating a linear regime. At higher drive powers, the resonance frequency shifts to a lower value, indicating the onset of the nonlinear regime.
The observed nonlinearity can be quantitatively analyzed within the framework of quantum circuit theory, as we discuss next.

The fundamental mode of the single JJ-based resonator can be modeled as a weakly anharmonic oscillator described by a Duffing Hamiltonian.
To quantify the Kerr nonlinearity, the resonance frequency at each drive power is extracted by tracking the dip in the $S_\mathrm{11}$ magnitude response. The corresponding average photon number in the resonator is then calculated using equation $n = \frac{2}{\hbar \omega_\mathrm{r}^2}\frac{Q_\mathrm{L}^2}{Q_\mathrm{C}}P_\mathrm{in}$,
where $P_\mathrm{in}$ is the input power, $\omega_\mathrm{r}$ is the resonance frequency, and $Q_\mathrm{L}$ and $Q_\mathrm{C}$ are the loaded and coupling quality factors of the resonator, respectively~\cite{Ganjam2024, Gupta2025}. The dependence of the resonance frequency on photon number is subsequently analyzed using a linear relation $\omega(n)=\omega_0 + K n$, where $\omega_0$ is the low-power linear resonance frequency, $n$ is the average photon number, and $K/2\pi$ is the Kerr coefficient.
A linear regression of the measured resonance frequency as a function of photon number, shown in Figure \ref{fig:f2}d, yields the Kerr coefficient, with a small associated uncertainty. This analysis confirms the emergence of nonlinearity in the high-power regime and provides a quantitative measure of the nonlinear Kerr coefficient.

The direct galvanic coupling in our device enables DC biasing of the JJ and unique exploration of its nonlinearity, as we discuss next. The device response as a function of the applied DC bias is first characterized at a fixed gate voltage of $10$ V and a temperature of $20$ mK, as shown in Figure \ref{fig:f3}a. Here, the microwave probe power is set in the low-power linear regime. As the bias current increases, the resonance shifts to lower values and eventually disappears, consistent with the suppression of superconductivity and the switching of the junction into the normal state. The power-dependent response is further investigated at $20$ mK and a gate voltage of $10$ V for different bias currents. A representative dataset at a bias current of $1.5$~$\mathrm{\mu A}$ is shown in Figure \ref{fig:f3}b. For each bias current, the resonant frequency at a given signal power is extracted by tracking the dip in the magnitude of the reflection coefficient $|S_\mathrm{11}|^2$. These power-dependent frequency shifts are then used to determine the Kerr coefficient via fitting procedures performed at each bias current, as summarized in Figure \ref{fig:f3}c. An increase in the extracted Kerr coefficient with DC bias is observed, indicating enhanced nonlinearity of the junction. This behavior can be understood within the framework of the resistively and capacitively shunted junction (RCSJ) model, where an increased bias current leads to a greater tilt of the washboard potential, thereby modifying the curvature of the cosine potential wells and enhancing the nonlinearity experienced by the phase particle. A schematic representation of this mechanism is provided in Figure \ref{fig:f3}c inset. We model the JJ potential in the presence of a DC bias using the RCSJ picture given in the Supplementary Information section IV, and the calculated nonlinear coefficients are plotted in Figure \ref{fig:f3}d. The DC bias introduces a non-zero $c_3$ term, which is typically zero in single-JJ-based resonators without DC bias. With an increase in bias current, both the magnitudes of $c_3$ and $c_4$ increase, reflecting increased nonlinearity in the system and qualitatively capturing our experimental trend. Additional data from another device showing similar trends is provided in the Supplementary Information section V.

Analogous to the analysis of Kerr nonlinearity as a function of DC bias, we extract the Kerr coefficient at different gate voltages by performing fits to the power-dependent frequency shift at each electrostatic configuration. The resulting dependence of the Kerr coefficient on gate voltage is presented in Figure \ref{fig:f4}a. As the gate voltage is increased, the magnitude of the Kerr coefficient decreases, indicating a reduction in the effective nonlinearity. This trend can be attributed to increased junction transparency at higher gate voltages. As we discussed, for an SNS JJ with a high-transparency channel, the magnitude of the $c_4$ term decreases, translating to a decrease in nonlinearity. At higher gate voltages, the Kerr coefficient tends to saturate. Our observation of gate dependence of the Kerr nonlinearity is consistent with a previous report on a gate-tunable semiconductor JJ-based device~\cite{phan2023}.

For high-temperature operation of graphene-based JPAs and single-photon sensors, the stability of nonlinearity at elevated temperatures is important. Hence, we have also studied the temperature dependence of the Kerr coefficient, as shown in Figure \ref{fig:f4}b. In our devices, we observe significant nonlinearity even at $\sim1$ K, which will be useful in designing amplifiers and sensors for elevated-temperature operations. A monotonic reduction in the magnitude of the Kerr coefficient is observed with increasing temperature. This behavior can be attributed to a decrease in the magnitude of $c_4$, which weakens the nonlinearity. The analytically calculated temperature dependence of $c_4$ is provided in the Supplementary Information section IV. In Table 1, we compare the Kerr nonlinearity across various JJ-based devices. We observe large nonlinearity in our devices due to the high participation of the junction inductance, resulting from the lumped-element implementation of the LC resonator compared to transmission-line-based designs.

In summary, we have performed a comprehensive study of the tunable Kerr nonlinearity in graphene JJs. The nonlinearity is tunable via external knobs such as DC bias and electrostatic gating. We studied the effect of DC bias on the Kerr coefficient in detail and demonstrated the emergence of $c_3$ nonlinearity by making the potential asymmetric with bias current. In the future, the DC-bias-driven JPA can be explored for parametric amplification with nonzero $c_3$ to implement a $3$-wave mixing scheme, similar to asymmetric SQUID-based JPAs with improved saturation powers~\cite{Frattini2017ThreeWave, Frattini2018SNAIL}. Taken together, our work will be a useful resource for future studies, enabling the design of hybrid quantum devices, including qubits, amplifiers, and single-photon sensors. 

\section*{Acknowledgements}
We thank Shyam Shankar and R. Vijay for helpful discussions. We thank Supriya Mandal, Surat Layek, and Gaurav Sharma for helpful feedback on the manuscript. We acknowledge support from the Air Force Office of Scientific Research under award number FA2386-25-1-4027, the Department of Atomic Energy of the Government of India 12-R\&D-TFR-5.10-0100, and J.C. Bose Fellowship JCB/2022/000045 from the Department of Science and Technology of India. The preparation of hBN single crystals is supported by the Elemental Strategy Initiative conducted by MEXT, Japan (Grant Number JPMXP0112101001) and JSPS KAKENHI (Grant Numbers 19H05790 and JP20H00354).

\section*{Data Availability}
The experimental data used in the main text figures are available from the corresponding authors upon reasonable request.

\onecolumngrid
\setlength{\tabcolsep}{12pt} 
\begin{table*}[t!]
\centering
\begin{tabular}{|l | c | c | c | c |}
\hline
& & & & \\
Material platform & $K/2\pi$ (kHz) & $p$ & $\omega_0/2\pi$ (GHz) & $K/\omega_0$ \\
& & & & \\
\hline
& & & &\\
Al-AlO$_\mathrm{x}$-Al JJ based JPA~\cite{Bourassa2012,Boutin2017} & $-500$ & $\sim0.6$ & $4.95$ & $-101 \times 10^{-6}$ \\
& & & &\\

Graphene JJ based JPA~\cite{butseraen2022} 
& $-111$ & low & $5.885$ & $-18.86 \times 10^{-6}$ \\
& & & &\\

Al-InAs-Al JJ based JPA~\cite{hao2024} 
& $-3.30 \pm 1.00$ & 0.29 & $7.52$ & $-0.4388 \times 10^{-6}$ \\
& $-0.13 \pm 0.04$ & 0.11 & $6.63$ & $-0.019 \times 10^{-6}$ \\
& & & &\\
Our graphene JJ devices 
& $-467.38 \pm 2.21$ & large & $6.75$ & $-69.24 \times 10^{-6}$ \\
& $-356.10 \pm 0.70$ & ($\sim 0.9$) & $4.81$ & $-74.03 \times 10^{-6}$ \\
& & & &\\
\hline
\end{tabular}
\centering
\vspace{10pt}
\caption{Comparison of the Kerr coefficient across literature among different JJ systems, including Al-AlO$_\mathrm{x}$-Al, graphene, and semiconductor-based platforms. The $\omega_0$ is the linear resonance frequency of the resonator, and $p$ is the inductive participation ratio of the JJ.}
\label{tab:kerr}
\end{table*}
\twocolumngrid

\bibliography{bibli}

\onecolumngrid
\clearpage
\phantomsection
\begin{center}
    \textbf{\LARGE Supplementary Information}
\end{center}
\vspace{0.3in}
\twocolumngrid

\setcounter{section}{0}
\renewcommand{\thesection}{S\arabic{section}}
\renewcommand{\theHsection}{SI.\arabic{section}}

\setcounter{figure}{0}
\renewcommand{\thefigure}{S\arabic{figure}}
\renewcommand{\theHfigure}{SI.\arabic{figure}}

\setcounter{table}{0}
\renewcommand{\thetable}{S\arabic{table}}
\renewcommand{\theHtable}{SI.\arabic{table}}

\setcounter{equation}{0}
\renewcommand{\theequation}{S\arabic{equation}}
\renewcommand{\theHequation}{SI.\arabic{equation}}

\section{Device Fabrication}
Josephson junctions (JJs) are used in circuit quantum electrodynamics (cQED) to realize nonlinear LC resonators. For our device geometry, we employ a coplanar waveguide (CPW)–based, directly coupled design, where the central conductor is terminated to the ground plane through a lumped-element LC resonator (see Supplementary Figure \ref{fig:sif1cpw}). The device's nonlinear response and stability are primarily governed by the Josephson inductance participation ratio, $p = L_\mathrm{J} / L_{\mathrm{tot}}$, where $L_\mathrm{J}$ is the inductance of the Josephson junction (also known as the Josephson inductance) and $L_{\mathrm{tot}}$ includes all inductive contributions, including the unwanted stray inductances that may appear due to practical constaints of the geometry. The lumped-element CPW design enables $p \rightarrow 1$, thereby maximizing the effective nonlinearity. In contrast, distributed resonator architectures exhibit reduced participation due to the dominant contribution of transmission-line inductance, which weakens the nonlinear response. The whole device fabrication is performed in the following four major steps.

\begin{figure}[h!]
    \centering
    \includegraphics[width=0.98\linewidth]{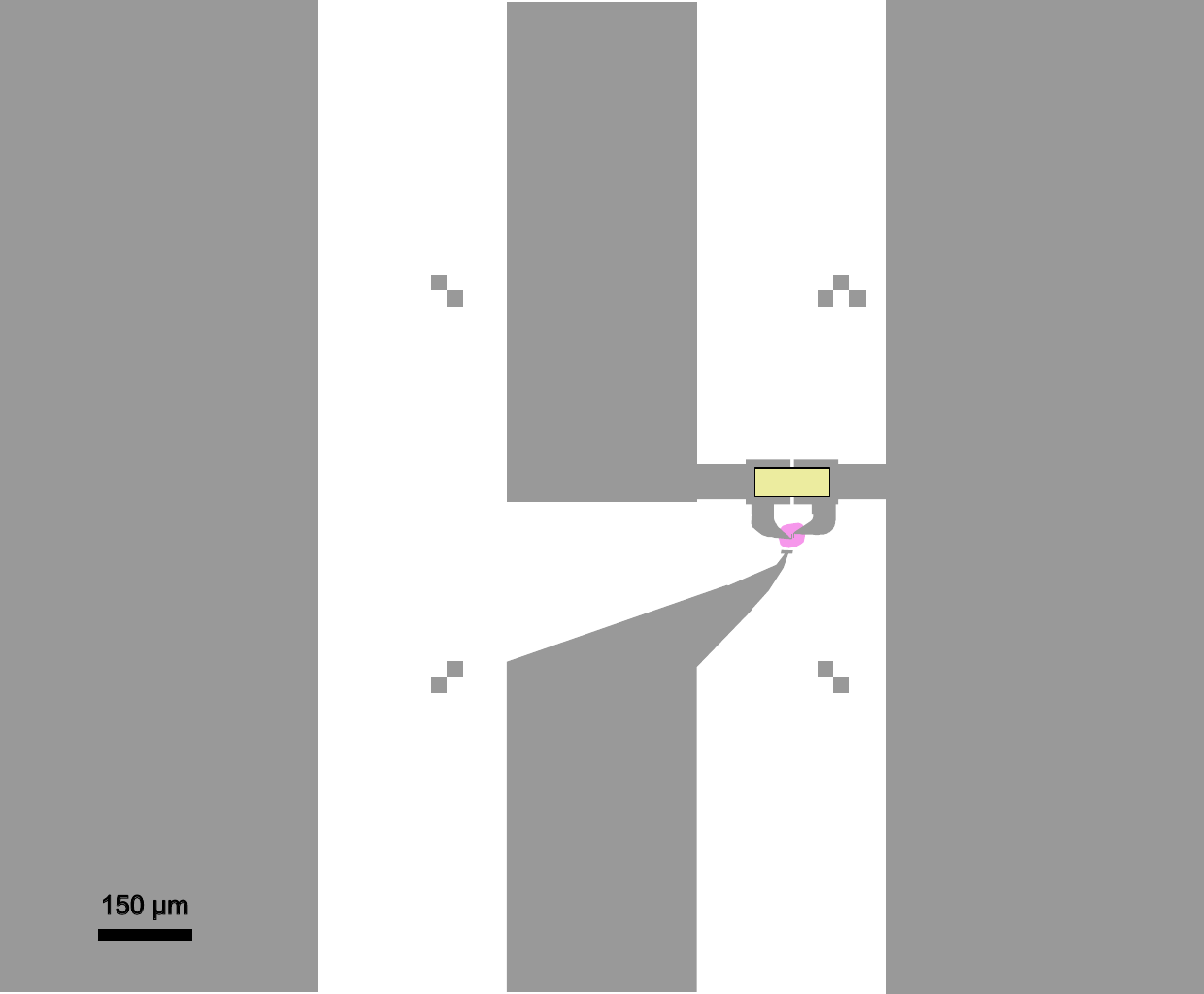}
    \caption{Shows the schematic representation of the coplanar wave guide geometry-based resonator. The central conductor and ground planes are designed to have a characteristic impedance of $50$ Ohm.}
    \label{fig:sif1cpw}
\end{figure}

\subsection{CPW}
The device fabrication proceeds in four main stages. First, a coplanar waveguide (CPW) is defined on a SiO$_2$/Si substrate using standard electron-beam lithography, followed by deposition of MoRe via DC magnetron sputtering under high-vacuum conditions. The CPW is designed to achieve a characteristic impedance of $50~\Omega$.

\subsection{Parallel plate capacitors}
In the second step, two MoRe–Al$_2$O$_3$–Al parallel-plate capacitors are fabricated. The dielectric and top metal layers are deposited using angled, rotational electron-beam evaporation to ensure uniform coverage. The Al$_2$O$_3$ dielectric thickness is $50$ nm, and it has a dielectric constant $\epsilon_r\sim9$. There are two parallel-plate capacitors in series, of identical dimensions $ 60$×$60$ $\mu$m$^2$, fabricated together for ease of fabrication, so that the Al$_2$O$_3$ and Al layers can be formed in the same step.

\subsection{hBN-graphene-hBN heterostructure}
Thin flakes of graphene and hexagonal boron nitride (hBN) are obtained via mechanical exfoliation of bulk crystals and subsequently stacked using a polymer (PC/PDMS)-based dry transfer technique. Few-layer graphene is selected and encapsulated between hBN flakes to make the hBN-graphene-hBN stack. Finally, a graphite flake is picked up, serving as a local back gate. The hBN–graphene–hBN-graphite stack is now dropped on the CPW substrate to make the JJ contacts and form the LC resonator. 

\subsection{Graphene Josephson junction}
Among the remaining steps, first, electron-beam lithography is done using PMMA 495 A4 and 950 A4 bilayer resist. Secondly, reactive-ion etching (CHF$_3$/O$_2$) is done to define the edge-contact trenches. Then, MoRe sputtering is performed to contact the superconducting electrodes with the graphene, forming the Josephson junction. Prior to MoRe deposition, an in situ Ar plasma cleaning step is performed to improve contact transparency. We deposit $40$ nm of MoRe side contacts and perform lift-off. To further improve contact transparency, the device is vacuum-annealed at $350$\textdegree C for $2$ hr in a $15$ sccm forming gas (N$_2$+H$_2$) environment. The fabricated chip is finally mounted on a printed circuit board and wire-bonded for electrical measurement. 

\section{Measurement setup}
\begin{figure}[ht]
    \centering
    \includegraphics[width=0.94\linewidth]{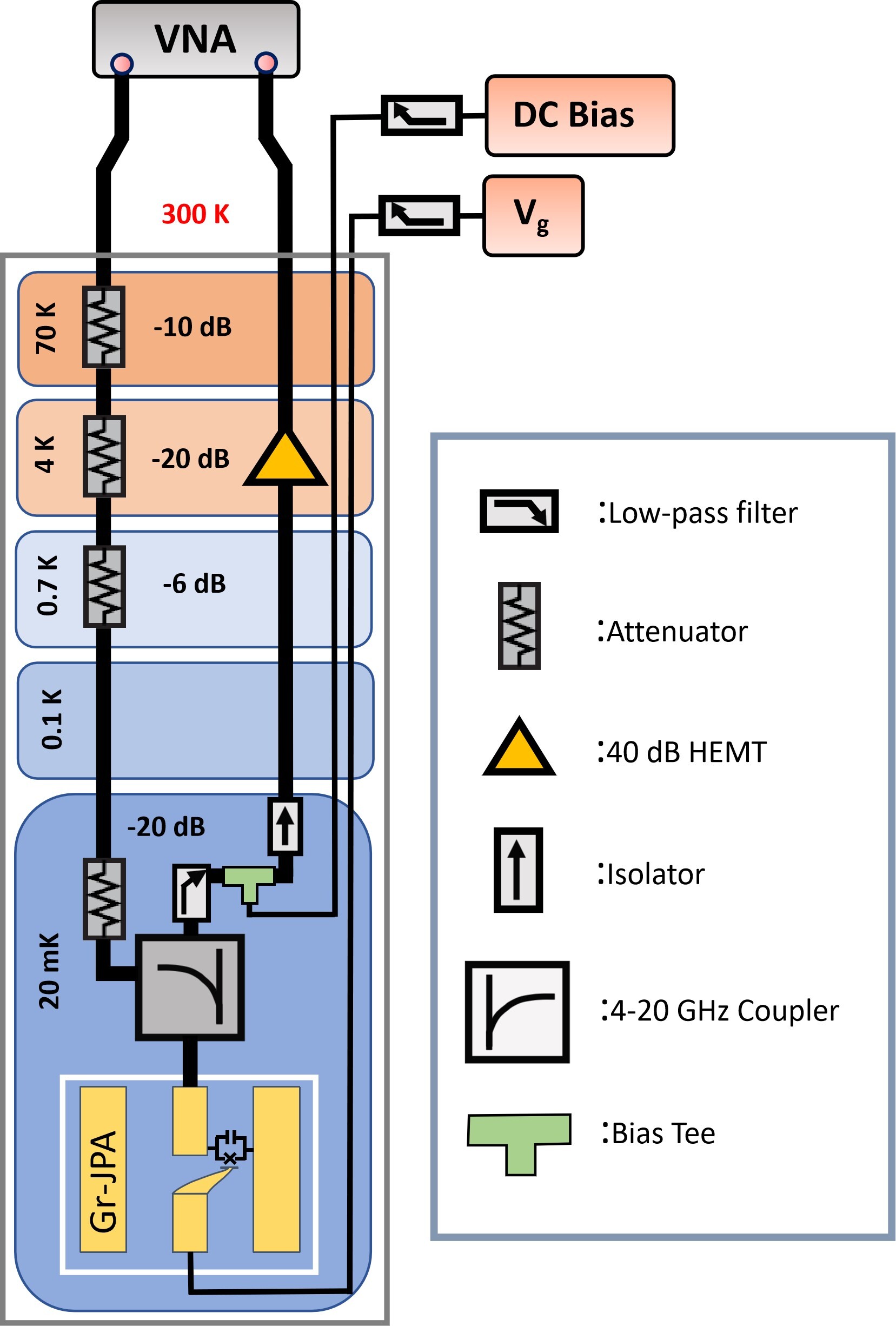}
    \caption{Wiring diagram and the microwave measurements setup. Shows the wiring diagram for the Kerr nonlinearity measurements.}
    \label{fig:sif2meas}
\end{figure}
The microwave measurements employ standard reflectometry in an Oxford dilution refrigerator at $20$ mK. 
During measurements, we take precautions to filter any undesired noise that may reach the device. The microwave setup, along with the wiring diagram, is shown in the Supplementary Figure \ref{fig:sif2meas}. The microwave lines are made of stainless-steel rigid coaxial lines. The input lines are attenuated by 56 dB using fixed attenuators at different temperature stages.  The output signal from the device passes through a directional coupler (Krytar $4-20$  GHz), a $10$ GHz low-pass filter, and a dual isolator (LNF) at the $20$ mK plate. The output signal is amplified by a 40 dB high-electron-mobility transistor (HEMT) amplifier at the $4$ K plate. The signal from the VNA (R \& S ZNB $20$) is sent to the device from port-$1$. The output signal is returned to the VNA at port-$2$. The gate voltage ($V_{\mathrm g}$) is applied to the device using a DC voltage source (NI-DAQ). The DC line for the gate passes through a $10$ Hz low-pass RC filter, followed by three-stage filtering using low-pass RC filters on different temperature plates of our dilution fridge (room temperature, $4$ K plate, $20$ mK plate), each with a cut-off frequency of 70 kHz. The DC-bias signal also passes through the aforementioned three-stage RC filters. The device is loaded into a copper PCB box with an aluminum puck for cryogenic electromagnetic shielding. 

\section{Derivation of Kerr nonlinearity expression from the JJ potential}
Now, we derive an analytic expression for the Kerr nonlinearity. The Kerr coefficient can be extracted from the power-dependent frequency shift by analyzing the equation of motion of a nonlinear Duffing oscillator. If we consider the phase difference across the JJ as $\phi$ and expand the potential energy around a stable minima point $\phi_0$. Writing $\delta \phi = \phi - \phi_0$, the Josephson potential can be expanded as
\begin{equation}
\begin{split}
U(\phi) &= U(\phi_0)
+ c_1 (\delta \phi)
+ \frac{1}{2!} c_2 (\delta \phi)^2
+ \frac{1}{3!} c_3 (\delta \phi)^3\\
&+ \frac{1}{4!} c_4 (\delta \phi)^4
+ \mathcal{O}((\delta\phi)^5),
\end{split}
\end{equation}
where $c_i$'s are the expansion coefficients. For a current-biased JJ, the RCSJ model gives the dynamic equation of motion of the phase particle, which is given by~\cite{tinkham1996introduction}
\begin{equation}
I_\mathrm{b} = I_\mathrm{s}(\phi) + \frac{\hbar}{2eR}\dot{\phi} + \frac{\hbar C}{2e}\ddot{\phi},
\end{equation}
where $I_\mathrm{b}$ is the bias current, $I_s(\phi)$ is the supercurrent flowing throgh the JJ, $R$ is the shunt resistance, and $C$ is the junction capacitance. Equivalently, this can be written in a mechanical form using the Josephson potential: $I_\mathrm{s}(\phi) = \frac{2e}{\hbar}\,\frac{\partial U}{\partial\phi}$ such that,
\begin{equation}
m_\phi \ddot{\phi} + \gamma_\phi \dot{\phi} + \frac{\partial U}{\partial \phi} = 0,
\end{equation}
where $U$ is the effective potential and the coefficients $m_\phi = \left(\frac{\hbar}{2e}\right)^2 C$ and  $\gamma_\phi = \left(\frac{\hbar}{2e}\right)^2 \frac{1}{R}$. Now expanding around the minima point $\phi_0$, the equation of motion becomes
\begin{equation}
m_\phi \ddot{\delta\phi}
+ \gamma_\phi \dot{\delta\phi}
+ c_2 \delta\phi
+ \frac{1}{2} c_3 (\delta\phi)^2
+ \frac{1}{6} c_4 (\delta\phi)^3
= 0.
\end{equation}
Neglecting damping for the purpose of extracting the frequency shift, the equation reduces to
\begin{equation}
\ddot{\delta\phi} + \omega_0^2 \delta\phi + \alpha (\delta\phi)^2 + \beta (\delta\phi)^3 = 0,
\end{equation}
where $\omega_0 = \sqrt{\frac{c_2}{m_\phi}}$ is analogous to the natural oscillation frequency of the phase particle, $\alpha = \frac{c_3}{2m_\phi}$, and $\beta = \frac{c_4}{6m_\phi}$ are the nonlinear coefficients of a Duffing oscillator. Now using the Lindstedt-Poincar\'e method, the amplitude-dependent frequency is given by~\cite{landau1976mechanics, nayfeh1981perturbation}
\begin{equation}
\omega(A)
=
\omega_0
+
\left(
\frac{3\beta}{8\omega_0}
-
\frac{5\alpha^2}{12\omega_0^3}
\right) A^2
+ \mathcal{O}(A^4),
\end{equation}
where $A$ is the oscillation amplitude of the phase variable $\delta\phi$ and $A^2$ represents drive power. This equation represents the drive-power-dependent frequency shift in a Duffing oscillator. Hence,
substituting $\alpha = \frac{c_3}{2m_\phi}$, $\beta = \frac{c_4}{6m_\phi}$ and $\omega_0^2 = c_2/m_\phi$ we get
\begin{equation}
\omega(A) = \omega_0 + \frac{A^2}{16 m_\phi \omega_0} \left(c_4 - \frac{5c_3^2}{3c_2} \right) + \mathcal{O}(A^4).
\end{equation}
Thus, the Kerr or effective nonlinear coefficient is
\begin{equation}
\boxed{K \propto c_4 - \frac{5c_3^2}{3c_2}}.
\end{equation}
For a symmetric potential, the $c_3$ term is zero, and the Kerr is set by the $c_4$. However, with DC biasing, the potential can be tilted, making it asymmetric and leading to a non-zero $c_3$ term. Another way to create a non-zero $c_3$ is to engineer the JJ potential, as done in asymmetric SQUID-based amplifiers ~\cite{Frattini2017}.

\begin{figure}[!ht]
    \centering
    \includegraphics[width=0.98\linewidth]{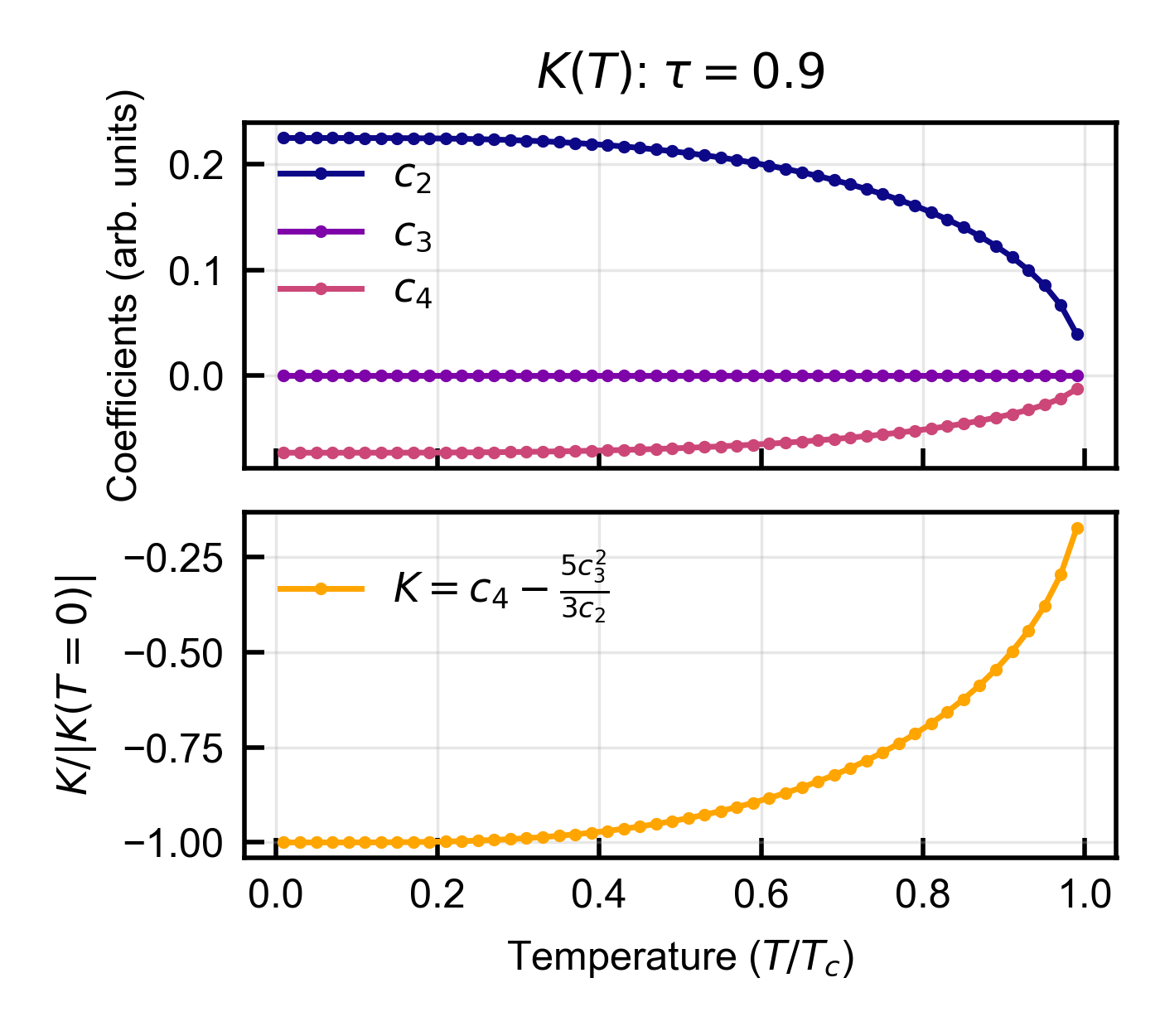}
    \caption{Temperature dependence of the non-linear coefficients is plotted. The coefficients are calculated by expanding the potential around the minima for an SNS junction with effective transparency $\tau=0.9$.}
    \label{fig:Temp}
\end{figure}

\section{Kerr nonlinearity with DC bias and temperature}

In this section, we discuss the theoretical modeling of the Kerr nonlinearity under current bias within the RCSJ picture. The Josephson potential for a DC-biased SNS JJ is given by
\begin{equation}
U(\phi, I_{DC}) = -E_J\left(\sqrt{1-\tau \sin^2(\phi/2})+\frac{I_{DC}}{I_c}\phi\right),
\end{equation}
where $\phi$ is the superconducting phase difference across the junction, $E_J$ is the Josephson energy, $\tau$ is the effective transparency of the SNS channel, $I_{DC}$ is the applied DC bias, and $I_c$ is the critical current of the JJ. The first term is the Josephson potential arising from the formation of Andreev bound states, while the second term arises from the tilt of the potential induced by the applied bias current. The analytically calculated coefficients $c_2$, $c_3$, and $c_4$ are plotted in the main text Figure 3d. The DC bias tilts the potential, making it asymmetric and leading to a non-zero $c_3$ coefficient that increases with increasing DC bias. Similarly, the junction potential as a function of temperature in a ballistic SNS JJ in the $T\ll T_c$ limit can be written as 
\begin{equation}
    U(\phi, T) = -\Delta(T) \sqrt{1 - \tau \sin^2(\phi/2)},
\end{equation}
where $\Delta(T) = \Delta_0 \tanh\!\left(1.74 \sqrt{T_c/T - 1}\right)$ is the temperature dependence of the superconducting energy gap according to BCS theory, $\Delta_0$ is the superconducting energy gap in the zero temperature limit, $T_c$ is the critical temperature of the superconducting leads, $T$ is the absolute temperature~\cite{Beenakker1991PRL, Furusaki1991PRB}. The temperature dependence of the non-linear coefficients is plotted in Supplementary Figure \ref{fig:Temp}. As the temperature increases, the magnitude of $c_4$ decreases, reducing the nonlinearity.

\section{Additional data from device D-1 and D-2}
Here we show additional Kerr coefficient vs. DC bias data at different gate voltages for D-1 (see Supplementary Figure \ref{fig:D1D2}), as well as data from another device (D-2) (see Supplementary Figure \ref{fig:D2}).

\begin{figure}[!ht]
    \centering
    \includegraphics[width=0.94\linewidth]{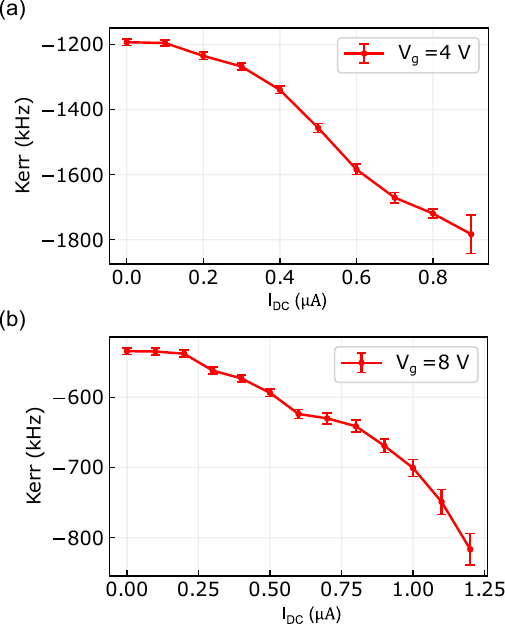}
    \caption{Additional data from the first device D-1: Fixing at particular gate voltages, Kerr nonlinearity extracted from the linear regression as discussed in the main text. \textbf{(a)} At $V_\mathrm{g}$= 4 V, the Kerr is extracted as a function of DC bias. \textbf{(b)} At $V_\mathrm{g}$= 8 V, the Kerr is extracted as a function of DC bias.}
    \label{fig:D1D2}
\end{figure}

\begin{figure*}[!ht]
    \centering
    \includegraphics[width=0.9\linewidth]{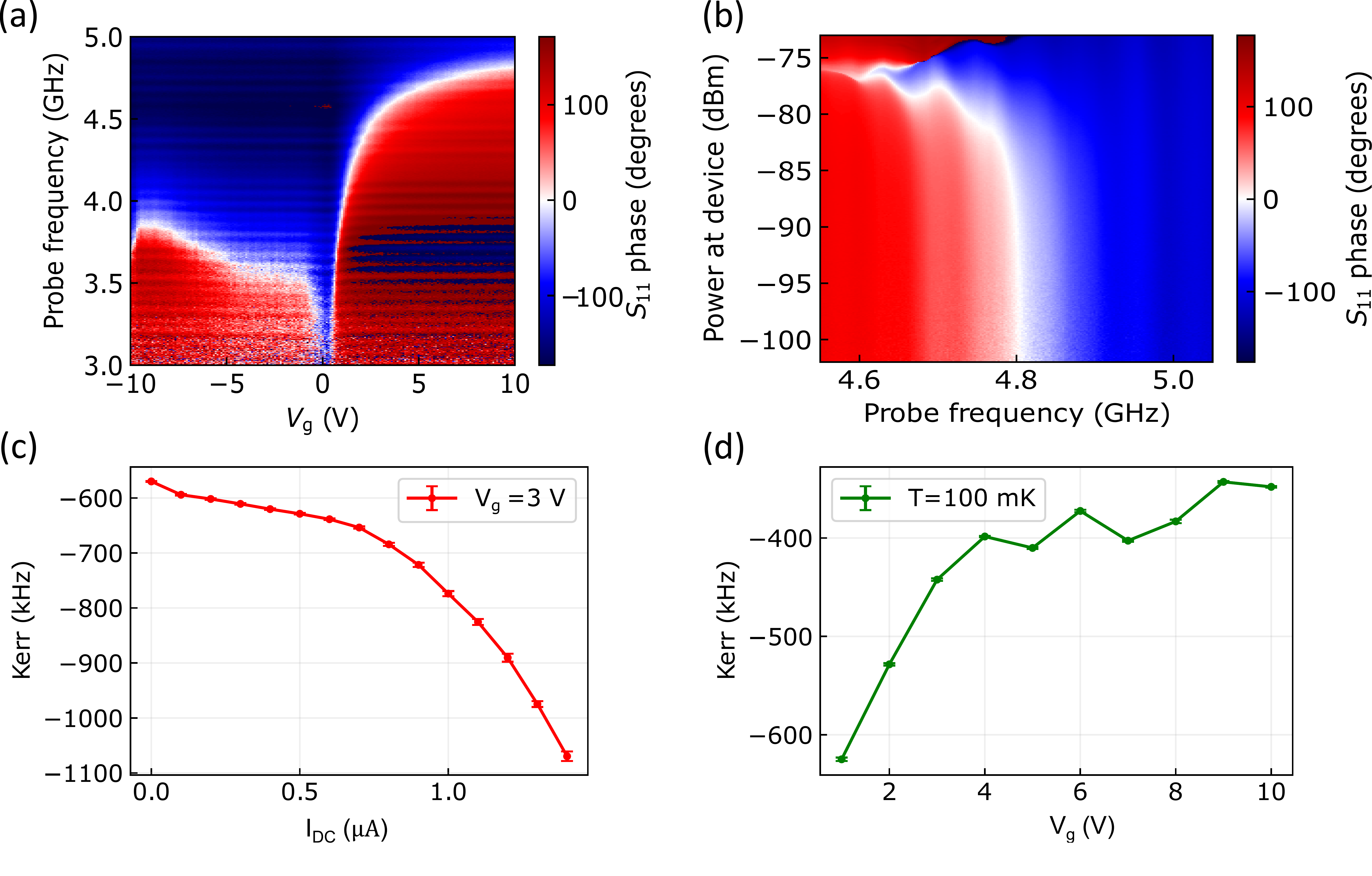}
    \caption{Additional data from the second device D-2: 
    \textbf{(a)} Shows the variation of resonance frequency with gating. The S$_{11}$ phase is plotted as a function of probe frequency and gate voltage.
    \textbf{(b)} The measured nonlinear response of the device. The resonance frequency remains constant in the low-power linear regime. At high powers, the resonance frequency shifts downward due to the nonlinearity. The data is taken at $V_\mathrm{g}=10$ V.
    \textbf{(c)} At $20$ mK, and $\mathrm{V_g}$ = 3 V, the Kerr is extracted with increasing DC bias.
    \textbf{(d)} At $100$ mK, zero DC bias, the Kerr is extracted as a function of gate voltage.}
    \label{fig:D2}
\end{figure*}

\newpage
\bibliographystyle{apsrev4-1}

\end{document}